\documentclass[pre,aps,onecolumn]{revtex4}

\usepackage{graphicx}
\usepackage{amsmath}
\usepackage{amssymb}
\usepackage{ifthen}
\usepackage{booktabs}
\usepackage{color}

\usepackage{graphicx}% Include figure files
\usepackage{dcolumn}% Align table columns on decimal point
\usepackage{bm}% bold math
\usepackage{longtable}
\usepackage{gensymb}

\newcommand{\bb}{\begin{equation}}
\newcommand{\ee}{\end{equation}}
\newcommand{\ba}{\begin{eqnarray*}}
\newcommand{\ea}{\end{eqnarray*}}
\newcommand{\rhor}{\rho({\bf r})}
\newcommand{\dd}{{\rm d}}
\newcommand{\rr}{{\mathbf r}}
\newcommand{\dr}{{\rm d}{\bf r}}

\usepackage{graphicx}% Include figure files
\usepackage{dcolumn}% Align table columns on decimal point
\usepackage{bm}% bold math
\usepackage{longtable}

\begin{document}

%\preprint{}

\title{Bridging transitions for spheres and cylinders}

\author{Alexandr \surname{Malijevsk\'y}}
\affiliation{ {Department of Physical Chemistry, Institute of
Chemical Technology, Prague, 166 28 Praha 6, Czech Republic;}\\
 {Laboratory of Aerosols Chemistry and Physics, Institute of Chemical Process Fundamentals, Academy of Sciences, 16502 Prague 6, Czech Republic}}
\author{Andrew O. \surname{Parry}}
\affiliation{Department of Mathematics, Imperial College London, London SW7 2BZ, UK}

\begin{abstract}
We study bridging transitions between spherically and cylindrically shaped particles (colloids) of radius $R$ separated by a distance $H$ that are dissolved in a bulk
fluid (solvent). Using macroscopics, microscopic density functional theory and finite-size scaling theory we study the location and order of the bridging transition and
also the stability of the liquid bridges which determines spinodal lines. The location of the bridging transitions is similar for cylinders and spheres, so that for
example, at bulk coexistence  the distance $H_b$ at which a transition between bridged and unbridged configurations occurs, is proportional to the colloid radius $R$.
However all other aspects,  and, in particular, the stability of liquid bridges, are very different in the two systems. Thus, for cylinders the bridging transition is
typically strongly first-order, while for spheres it may be first-order, critical or rounded as determined by a critical radius $R_c$. The influence of thick wetting
films and fluctuation effects beyond mean-field are also discussed in depth.
\end{abstract}

%Use showkeys class option if keyword

%\maketitle must follow title, authors, abstract, \pacs, and %\keywords
\maketitle

\section{Introduction}

It is well known that effective pair interactions between dissolved particles may be induced by the presence of the solvent. A much studied example of this phenomenon is
a solvent-mediated depletion interaction between colloids which occurs when their size is substantially larger than that of the solvent particles \cite{lek}. In this
case, the entropy gain due to excluded volume give rise to an effective short-ranged attraction between the colloids determined by the diameter of the solvent. The force
of depletion may then lead to clustering of the solute particles and plays a crucial role in the stability of colloidal and colloidal-polymer systems.

Another type of effective force may occur when the solvent forms a liquid or gas bridge between the colloidal particles. The details of this depend sensitively on the
wetting properties of the colloidal particles, the distance between them and also how close the solvent is to its own fluid-fluid phase boundary. While such bridging
induced interactions require finer thermodynamic tuning than the force of depletion, they are stronger and also longer-ranged since they are determined by the size of
the colloidal as opposed to solvent particle. Bridging between colloidal spheres was originally studied using macroscopic arguments \cite{macro_rings, macro_vogel,
macro_willet} and simple effective interfacial models \cite{yeomans, yeomans2}. In such macro/mesoscopic models the transition between unbridged and bridged
configurations must be first-order owing to the different symmetries of the interfacial profiles. However, more microscopic approaches based on density functional theory
(DFT) have shown that the scenario is more complex \cite{dietrich, archer1, archer2, mal}. In particular,  as shown recently \cite{mal}, the phenomenon is much more
similar to capillary-condensation and criticality in parallel-plate geometries. This means that for spherical colloids the formation of a liquid bridge only occurs via a
first-order transition for sufficiently large sphere radius $R>R_c$. Surprisingly for $R<R_c$ the formation of a bridge is continuous and occurs via the smooth merging
of the wetting layers surrounding each colloid -- something which is beyond the reach of mesoscopic models.

In this paper we compare and contrast the character of the bridging transition between spheres and between parallel aligned cylinders  to which much less attention has
been devoted \cite{macro_cyl}. We show that there are major quantitative differences between these two types of bridging which are apparent at macroscopic level, within
microscopic DFT and beyond mean-field when fluctuation effects are considered. Of particular importance is the stability of liquid bridges, which are very different for
the spheres and cylinders, both at and off bulk coexistence. Thus for example at bulk coexistence for cylinders our DFT studies indicate that, at least away from the
immediate vicinity of the bulk critical point, and in contrast to the case of spheres, there is no critical radius $R_c$, so that a first-order bridging transition is
always present. We show that at two phase coexistence, the shapes of the minimal surfaces forming the bridge lead to universal results for the location of the bridging
transition for large particle sizes. This is closely analogous to the Kelvin equation for condensation in capillary slit geometries and implies, in the macroscopic
limit, that the distance $H$ between the particles at which bridging occurs scales linearly with the particle size $R$. Corrections to this resulting from the presence
of thick complete wetting layers are also predicted and tested using DFT, as is the case when the solvent fluid is off two-phase coexistence for which a non-linear phase
boundary and spinodal are constructed. The finite-size rounding of the bridging transition due to fluctuation effects is also discussed.

\section{Macroscopic theory}

In this section, we formulate a simple macroscopic theory for bridging transitions between two identical parallel cylinders and between a pair of identical spheres. The
distance between these, measured centre-to-centre (for spheres) and axis-to-axis (for cylinders), is $H$ while the radius in each case is $R$. These are immersed in a
vapour at temperature $T<T_c$ with $T_c$ the critical temperature and chemical potential $\mu$ which may be at or away from bulk two phase coexistence $\mu=\mu_{\rm
sat}(T)$. For the most part we will assume that the spheres/cylinders are completely wet by liquid so that the contact angle for each is zero although it is
straightforward to generalise the analysis to partial wetting which is also discussed briefly. If the spheres/cylinders are far apart they are each coated with a layer
of liquid the thickness of which is determined by $R$ and the range of the intermolecular forces. However, the forces and the thickness of the wetting layers need not be
specified in the macroscopic regime. For smaller separations $H$ on the other hand it is possible that liquid condenses in the region between the spheres/cylinders
forming a bridge between them. This reduces the area exposed by the wall-gas interface but increases that of the wall-liquid and liquid-gas interface. A free-energy
balance between these two configurations occurs at $H=H_b(R)$, determined by the radius $R$, corresponding to the macroscopic bridging transition. At this level the
transition is always first-order since the configurations are always distinct. However, the line of the bridging coexistence curves $H_b(R)$ is different for the
cylinders and the spheres because the shapes of the liquid bridges, which are minimal surfaces at bulk coexistence, have very different properties. In the following, let
$\gamma_{\rm wl}$, $\gamma_{\rm wg}$ and $\gamma_{\rm lg}$ represent surface tensions between wall-liquid, wall-gas and liquid-gas interfaces, respectively, which are
assumed to be the same as those of the planar surfaces. The differences between the free energies of the unbridged and bridged configurations arise simply from the
surface tensions and the appropriate area exposed. We consider this for cylinders and spheres separately.

\subsection{Bridging between cylinders} \label{mt_cyl}

\begin{figure}
\includegraphics[width=0.5\textwidth]{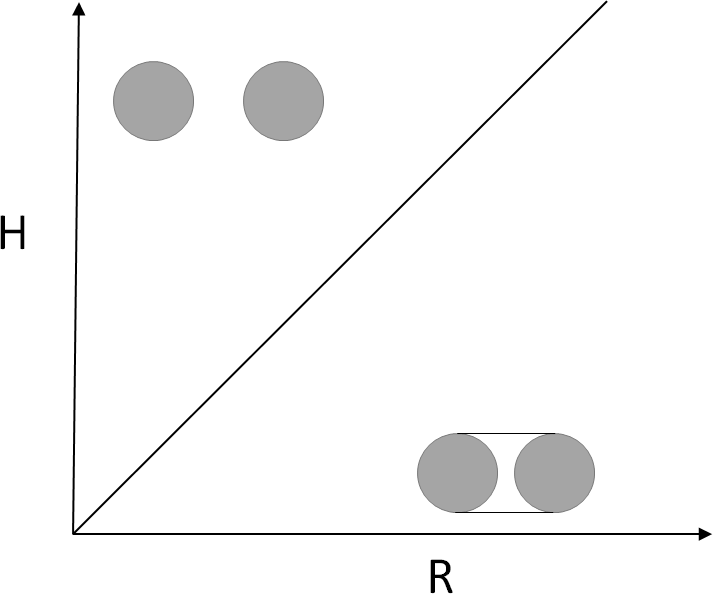}
\caption{Macroscopic phase diagram for first-order bridging transitions between completely wet cylinders immersed in a solvent at two-phase coexistence. The line
represents the bridging coexistence curve $H_b=\pi R$. The bridged phase remains metastable for all $H>H_b$.  } \label{fig1}
\end{figure}

\begin{figure}
\includegraphics[width=0.5\textwidth]{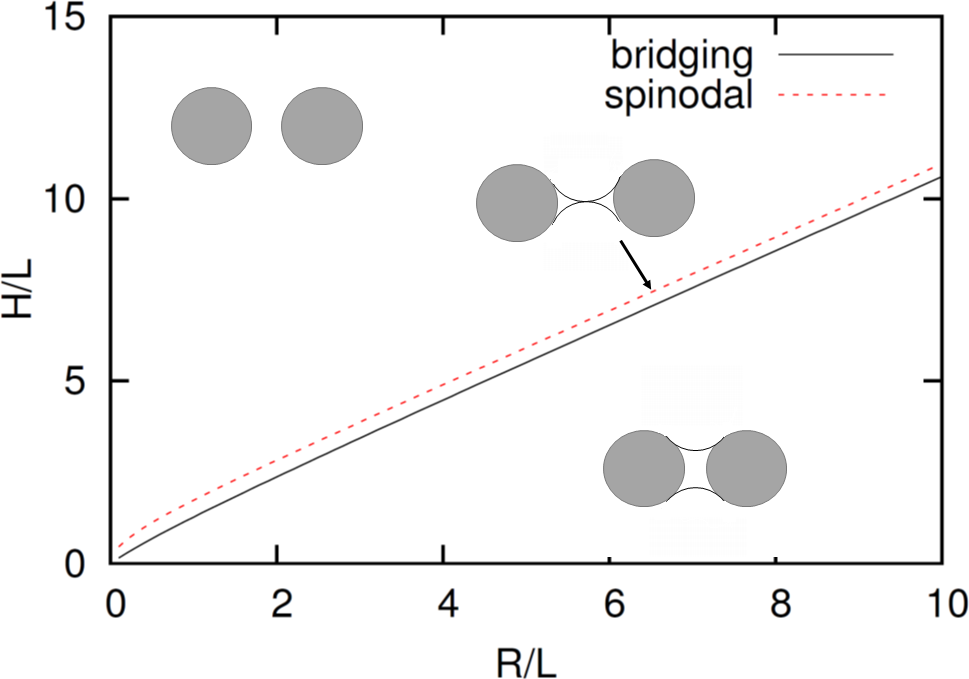}
\caption{ (Color online) Macroscopic phase diagram for first-order bridging transitions between completely wet cylinders immersed in a solvent off coexistence. The solid
line represents the bridging coexistence curve $H_b/L=2R/Lf(R/L)$ whilst the dashed line is the spinodal curve $H_{\rm spin}/L=2\sqrt{\frac{R^2}{L^2}+\frac{2R}{L}}$ at
which the bridging film pinches. } \label{fig2}
\end{figure}

%\begin{figure}
%\includegraphics[width=0.5\textwidth]{cyl_unb}
%\includegraphics[width=0.5\textwidth]{cyl_brd}
%\caption{A two-dimensional projection of  (a) unbridged and (b) bridged configurations of two infinitely long aligned cylinders. The system is translation invariant
%along the $y$-axis. The dashed line represents the macroscopic location of the liquid-gas interface.} \label{fig1}
%\end{figure}

Let us begin by considering the case of cylindrical colloids that are completely wet ($\theta=0$) and a solvent that is at bulk coexistence. In the unbridged state the
excess free-energy per unit length of the cylinder is $F=4\pi \gamma_{\rm wg}R$ coming from two separate cylinders. In the bridged configuration the cylinders are
connected by a liquid film consisting of two parallel flat liquid-gas interfaces. The curvature of these is trivially zero since we are at bulk coexistence. These flat
interfaces meet the cylinders tangentially since the contact angle is zero. The excess free-energy for this configuration is $F=2\pi R\gamma_{\rm wl}+2\pi R\gamma_{\rm
wg}+2H\gamma_{\rm lg}$, implying that the bridging transition occurs when
 \bb
 H_{b}=\pi R\label{kelvin_cyl}\,.
 \ee
This is equivalent to the Kelvin equation for condensation in capillary slits but now written in terms of the cylinder separation rather than the undersaturation. Note
that this simple linear relationship is to be expected from dimensional considerations since $R$ and $H$ are the only macroscopic lengths in the problem. The macroscopic
phase diagram for bridging transitions occurring at bulk coexistence is therefore particularly simple and is shown in Fig. 1. For fixed $R$ and $H>H_b$ the liquid bridge
is metastable with respect to the unbridged configuration, however there is no spinodal determining the limit of stability since flat interfaces may always span the two
cylinders.

The result is readily generalised to the case of partial wetting for which the contact angle $\theta>0$. The liquid-gas interfaces of the bridge phase are still flat but
now meet the cylinders at the contact angle rather than tangentially. A simple calculation yields
 \bb
 \frac{H_b}{R}=(\pi-2\theta)\cos\theta+2\sin\theta\,, \label{cos_cyl}
 \ee
which is symmetric about $\theta=\pi/2$. Notice that $H_b=2R$ for neutral wetting, $\theta=\pi/2$, which means that the macroscopic bridging transition no longer exists
as the cylinders are in contact. The value of $H_b$ is maximum for the case of complete wetting ($\theta=0$) or complete drying ($\theta=\pi$). For the latter scenario,
a bridge of gas forms between the cylinders when it is immersed in a  solvent of bulk liquid.

For the case of complete wetting one can also consider the  influence of the singular part of the wall-gas surface tension arising from complete wetting layers. For
example, for the case of short-range intermolecular forces, the wall-gas surface tension is modified to \cite{gelfand}
 \bb
 \gamma_{\rm wg}=\gamma_{\rm wl}+\gamma_{\rm lg}+\frac{\gamma_{\rm lg}}{R}\xi_l\ln\left(\frac{R}{\xi_l}\right)+\cdots\,,\label{cyl_derj}
 \ee
where $\xi_l$ is the bulk correlation length of the solvent in the liquid phase. Making use of this result, the bridging transition is shifted to
 \bb
 H_b\approx\pi R+\pi\xi_l\ln\left(\frac{R}{\xi_l}\right) \,,\label{kelvin_cyl2}
 \ee
implying that the wetting films enhance the onset of the transition by increasing the effective cylinders size. Similar corrections occur for dispersion-like forces
where they are algebraic of order $R^\frac{1}{3}$ \cite{bieker, stewart, nold}.  The same corrections occur for complete drying films. Thus for systems with short-ranged
forces, the expression for $H_b$ is of the form (\ref{kelvin_cyl2}) but with the bulk gas correlation length $\xi_g$ replacing $\xi_l$.

Finally let us return to the simple macroscopic theory for complete wetting and consider what happens off bulk coexistence when the chemical potential or pressure
departure from two-phase coexistence $\delta p=p_{\rm sat}-p$ is non-zero. In this case the liquid-gas interfaces forming the liquid bridging film are not flat but
rather arcs of circles of radius $L=\gamma_{\rm lg}/\delta p$ as required by Laplace's equation. The free-energy balance argument must now also include a term from the
cross-sectional area of the bridge which is proportional to $\delta p$ and a simple calculation shows that the transition occurs when
 \bb
 H_b(R,L)=2Rf\left(\frac{R}{L}\right)\,,\label{kelvin_cyl_off}
 \ee
 where $f(x)$ is a scaling function satisfying the implicit equation
 \bb
 \sin^{-1}\frac{xf}{x+1}=\frac{x}{(1+x)^2}\left(\frac{\pi}{2}(2+x)-f\sqrt{(1+x)^2-x^2f^2}\right)\,.\label{kelvin_cyl_off2}
 \ee
It is easy to show that $f(0)=\pi/2$ while $f(\infty)=1$ so that in the limit $R/L\to 0$ and $R/L\to \infty$ one obtains the linear results $H_b\approx \pi R$ and
$H_b\approx 2R$, respectively. The first limit therefore recovers our earlier result for bridging transitions at bulk coexistence. The second limit means that when the
colloid radius is larger than the Laplace radius, the bridging transition occurs very close to the contact distance. Interestingly, this generalised phase diagram now
includes a spinodal curve, see dashed line in Fig.~2, $H_s=2\sqrt{R^2+2LR}$, which marks the limits of stability of the bridging phase, corresponding to the pinching of
the bridge film in which the two arc circles of radius $L$ meet. Notice that the spinodal curve $H_b\approx \sqrt{8LR}$ for small $R/L$ so that the stability of the
bridge increases as coexistence is approached. However in the opposite limit, $R/L\to\infty$ the curves $H_s$ and $H_b$ have the same asymptote so that the stability of
the bridge is reduced. This means that when $R\gg L$, the bridging transition occurs near contact and is weakly first-order. This is the macroscopic mechanism for the
disappearance of the bridging transition as $T\to T_c$, since in that limit, $L\to 0$ due to vanishing of the surface tension $\gamma_{\rm lg}$.

\subsection{Bridging between spheres} \label{mt_sph}

%\begin{figure}
%\includegraphics[width=0.5\textwidth]{cat_unb}
%
%\vspace*{0.5cm}
%
%\includegraphics[width=0.5\textwidth]{cat_brd}
% \caption{A schematic illustration of  (a) unbridged and (b) bridged configurations for two spherical colloids.}
%\label{fig2}
%\end{figure}

\begin{figure}
\includegraphics[width=0.5\textwidth]{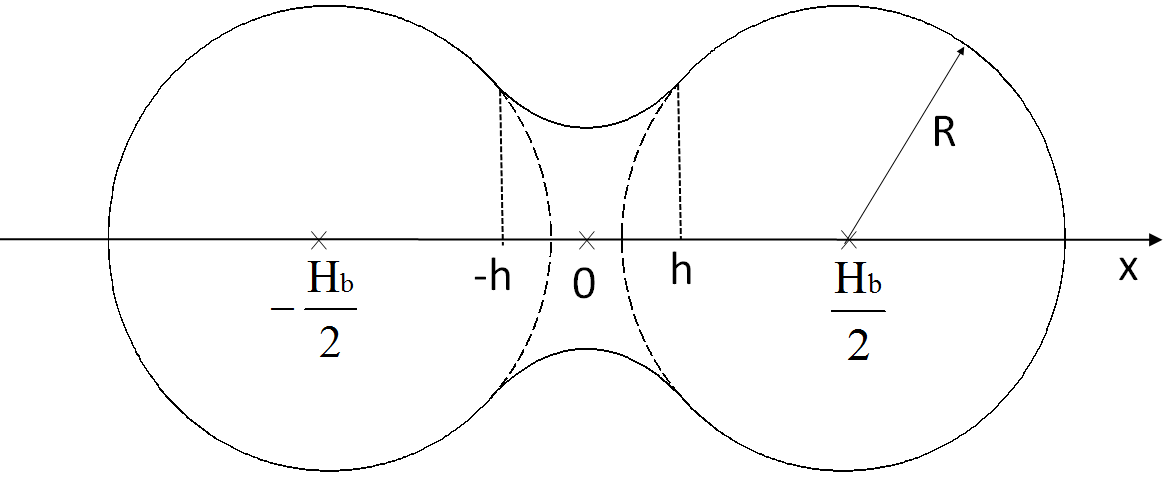}
 \caption{Illustration of the length-scales characterizing a macroscopic bridged configuration for spheres, see text for details.}
\label{sketch_sphere}
\end{figure}

\begin{figure}
\includegraphics[width=0.5\textwidth]{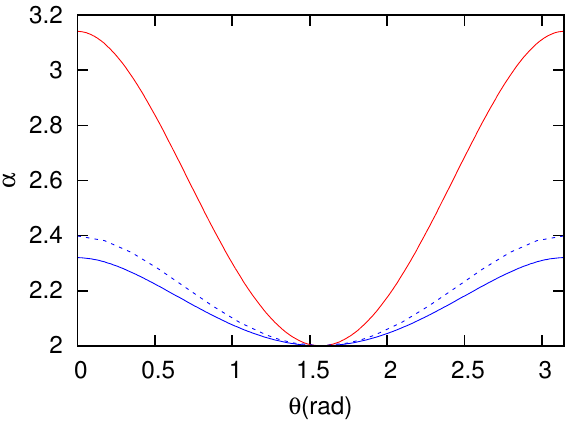}
\caption{(Color online) Contact angle dependence of the parameter $\alpha(\theta)$ appearing in the macroscopic prediction $H_b=\alpha R$  for spheres (blue) and
cylinders (red, uppermost) in which case the parameter $\alpha$ is given by Eq.~(\ref{cos_cyl}). No bridging is possible for $\theta=\frac{\pi}{2}$. The dashed line
represents the contact angle dependence of $\alpha_{\rm spin}(\theta)$ determining the spinodal $H_{\rm spin}=\alpha_{\rm spin}(R)$ for spheres which lies close to the
coexistence line.}\label{fig4}
\end{figure}

\begin{figure}
\includegraphics[width=0.5\textwidth]{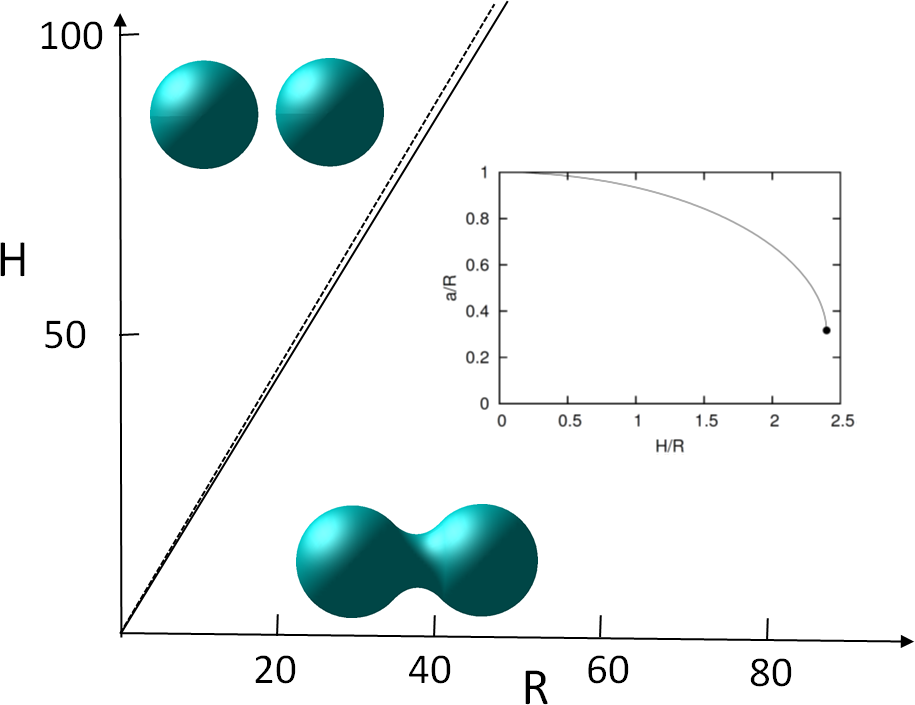}
\caption{(Color online) Macroscopic phase diagram for first-order bridging transitions between completely wet spheres immersed in a solvent at bulk coexistence. The
solid line represents the bridging coexistence curve $H_b\approx 2.32\,R$ whilst the dashed line is the spinodal curve $H_{\rm spin}/L\approx 2.38\,R$. The inset shows
the dependence of the parameter $a$, characterizing the catenoid shape $a\cosh\frac{x}{a}$, as a function of $H$. The termination of the curve determines the spinodal at
which the catenoid no longer touches the spheres. }\label{fig5}
\end{figure}

Next consider the bridging transition between two spherical colloids, each of a radius $R$, immersed in a bulk vapour at $\mu=\mu_{\rm sat}$. Due to the last condition,
the mean curvature of the macroscopic shape of the liquid bridge must be zero, i.e., must form a minimal surface. It is an easy task from a variational principle to show
that he minimal surface of a revolution  is a catenoid formed by rotating the catenary $y=a\cosh\frac{x}{a}$ about the revolution $x$-axis connecting the centres of the
spheres. The bridging transition occurs when there is a balance of surface free energies of the bridged and unbridged states balance equivalent to the condition
 \bb
 \frac{S_{\rm cat}}{S_{\rm wl}}=\frac{\gamma_{\rm wg}-\gamma_{\rm wl}}{\gamma_{\rm lg}}\label{df}
 \ee
where $S_{\rm cat}$ is the area of the catenoid and $S_{\rm wl}$ is the total area of the spheres covered by liquid in the bridged state. Following the illustration in
Fig.~\ref{sketch_sphere}, the catenoid connects with the spheres at $x=\pm h$ and therefore $S_{\rm wl}=4\pi R(R+h-H_b/2)$. The catenoid has area
 \bb
 S_{\rm cat}=2\pi a\int_0^h\cosh^2\left(\frac{x}{a}\right)dx=2\pi a\left[\frac{a}{2}\sinh\left(\frac{2h}{a}\right)+h\right]
 \ee
and substitution gives
 \bb
 \frac{a}{ R(2R+2h-H_b)}\left[\frac{a}{2}\sinh\left(\frac{2h}{a}\right)+h\right]=\cos\theta\,,\label{fb}
 \ee
where we have used Young's equation. The parameters $h$ and $a$ are determined from the contact conditions
 \bb
 a\cosh\left(\frac{h}{a}\right)=\sqrt{R^2-(h-H_b/2)^2} \label{con}
 \ee
 and
  \bb
  \tan^{-1}\left[{\frac{H_b/2-h}{\sqrt{R^2-(h-H_b/2)^2}}}\right]-\tan^{-1}\left[{\sinh\left(\frac{h}{a}\right)}\right]=\theta\,, \label{slope}
  \ee
where in the last equation we have allowed that the liquid-gas interface meets the spheres at an arbitrary contact angle. These equations have the solution
 \bb
 \frac{H_b}{R}=\alpha(\theta)\,,\label{kelvin_sph}
 \ee
where the function $\alpha(\theta)$ has to be determined numerically and is shown in Fig.~\ref{fig4}. The separation $H_b$ is largest for the case of complete wetting
(or complete drying), for which $\alpha(0)\approx2.32$ in an agreement with the analysis of \cite{dietrich}. The separation $H_b$ is smallest for neutral wetting
($\theta=\pi/2$) for which again $H_b=2R$ meaning that no bridging occurs. The linear relationship between $H_b$ and $R$ for $\delta p=0$ is similar to that found for
cylinders. However, the stability of the liquid bridge is completely different. In contrast to cylinders, the metastable region occurring for $H>H_b$ is of finite extent
and the spinodal line $H_{\rm spin}(R)$ lies close to the coexistence curve $H_b(R)$. The spinodal occurs not because the bridges pinch, as in the case of cylinders, but
because for large $H$ the catenoid has no solution touching the spheres. This is illustrated in Fig.~\ref{fig5} where we show the coexistence curve $H_b(R)\approx
2.32\,R$ and spinodal $H_{\rm spin}(R)\approx 2.38\,R$ for the case of complete wetting. In the inset we show the dependence of the length-scale $a$ characterising the
catenoid as a function of the sphere separation $H$. Notice that the value of $a/R=0.363$ at the spinodal is close to that at the bridging transition for which
$a/R=0.476$. More generally the spinodal line is always a linear function of the radius, $H_{\rm spin}(R)=\alpha_{\rm spin}(\theta) R$, but has a coefficient that
depends on the contact angle as shown in Fig.~\ref{fig4}. This shows that the extent of the metastable region is largest for the case of $\theta=0$ and smallest for
$\theta=\pi/2$. None of this behaviour is present for cylinders, where the spinodal line is only important off bulk coexistence. The close proximity of the spinodal line
for spheres is suggestive that the macroscopic bridging phase diagram at $\delta p=0$ may be more strongly influenced by microscopic effects, than for the phase diagram
for cylinders.

Finally we mention that similar to the cylindrical case, there are non-analytic corrections when complete wetting layers coat the spheres. For systems with short-ranged
forces these lead to
   \bb
 \frac{H_b}{R}\approx\alpha(0) +\frac{\alpha(0)}{2}\xi_l\ln\left(\frac{R}{\xi_l}\right) \,,\label{kelvin_sph2}
 \ee
in which the pre-factor of $2$ in the log correction arises from the curvature of the sphere. This means that the corrections due to complete wetting films are smaller
for spheres than for cylinders. Once again, for complete drying the bulk correlation function $\xi_l$ is replaced by $\xi_g$.

\section{Density functional theory}

Within classical density functional theory \cite{evans79} the equilibrium density profile $\rhor$ is found by minimising the grand potential functional
 \bb
 \Omega[\rho]={\cal{F}}[\rho]+\int\dr\left[V_i(\rr)-\mu\right]\rhor\,.\label{om}
 \ee
where $V_i(\rr)$, $i=s,c$, is the external field due to the pair of spherical or cylindrical colloids, respectively, and $\mu$ is the chemical potential. It is
convenient to split the intrinsic free energy functional ${\cal{F}}$ into the contribution due to the ideal gas and the remaining excess part
 \bb
 {\cal{F}}[\rho]={\cal{F}}_{\rm id}[\rho]+{\cal{F}}_{\rm ex}[\rho]\,,
 \ee
where ${\cal{F}}_{\rm id}[\rho]= k_BT\int\dr\rhor\left[\ln(\Lambda^3\rhor)-1\right]$ with $\Lambda$ the thermal de Broglie wavelength that can be set to unity.

Following the standard perturbative approach, the excess part of the free energy can be separated into repulsive and attractive contributions:
 \bb
{\cal{F}}_{\rm ex}[\rho]={\cal{F}}_{\rm HS}[\rho]+\frac{1}{2}\int\dr\rhor\int\dr'\rho(\rr')u_a(|\rr-\rr'|)\,,\label{fex}
 \ee
where $u_a(r)$ is the attractive part of the intermolecular potential.  In our calculations we take this to be that of a square-well (SW) potential
 \bb
 u_{\rm SW}=u_{\rm HS}+u_a\,,
 \ee
 where $u_{\rm HS}$ is the hard sphere potential with diameter $\sigma$ and
 \bb
 u_a(r)=\left\{\begin{array}{cc}
  0\,;& r\leq\sigma\,,\\
 -\varepsilon\,;& \sigma<r<\lambda\sigma\,,\\
 0\,;&r\geq\lambda\sigma\,,
\end{array}\right.
 \ee
where we set $\lambda=1.5$, as is usual.
%  \bb
% u(\tilde{r})=\left\{\begin{array}{cc}
% -\varepsilon\,;& \tilde{r}<\lambda\sigma\,,\\
% 0\,;&\tilde{r}\geq\lambda\sigma\,,
%\end{array}\right.
% \ee
%where $\tilde{r}=\sqrt{(z_2-z_1)^2+(r_2-r_1)^2}$ is the distance between the centres of two SW particles and $\lambda=1.5$, as usual.
The hard-sphere potential contribution to the free energy functional is included within  Rosenfeld's  fundamental measure theory \cite{rosenfeld89}:
 \bb
{\cal{F}}_{\rm HS}[\rho]=k_BT\int\dr\Phi(\{n_\alpha\})\,.\label{fhs}
 \ee
The free energy density $\Phi$ is a function of a set of three independent weighted densities $\{n_\alpha\}$ that can be expressed as two-dimensional integrals
\cite{mal, mal_grooves}. Here we adopt the original Rosenfeld prescription, although other forms of $\Phi$ are available \cite{mulero}.

In our tests of the macroscopic theory we will focus only on the case of complete drying ($\theta=\pi$) for spherical and cylindrical colloids in a bath of saturated
liquid. This is most easily modelled by using a purely hard-wall repulsion for the spheres and cylinders which induces drying at all temperatures $T<T_c$.

The minimisation of Eq.~(\ref{om}) leads to the Euler-Lagrange equation
 \bb
 k_BT\ln\Lambda^3\rho(x,r)+\frac{\delta F_{\rm ex}}{\delta\rho}+V_s(x,r)=\mu\,, \label{el1}
 \ee
 for spherical colloids (with $r$ the perpendicular distance to the $x$ axis) and
  \bb
k_BT\ln\Lambda^3\rho(x,z)+\frac{\delta F_{\rm ex}}{\delta\rho}+V_c(x,z)=\mu\,, \label{el2}
 \ee
for cylindrical colloids.

\begin{figure}
\includegraphics[width=0.5\textwidth]{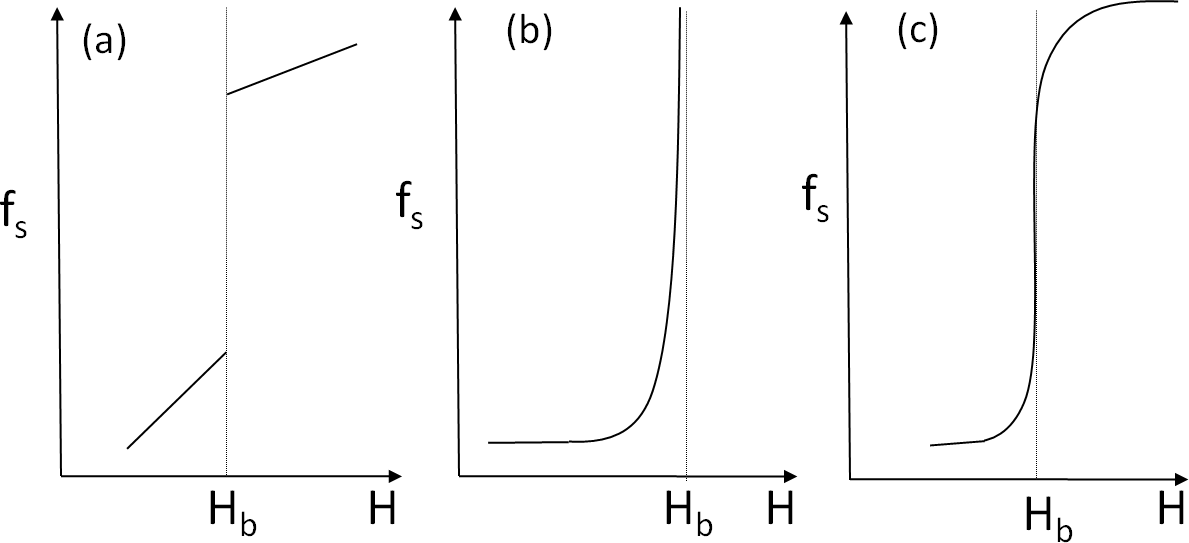}
\caption{Schematic behavior of the solvation force showing possible scenarios for bridging transition between spheres or cylinders: (a) first-order bridging, (b)
critical bridging and (c) rounded bridging.} \label{sketch_solv}
\end{figure}

The equations (\ref{el1}) and (\ref{el2}) are solved iteratively on a two-dimensional grid   with a uniform spacing of $0.05\sigma$ in all directions. We adopt a
Cartesian coordinate system for cylinders (exploiting a translation invariancy in the direction along the cylinders) and a cylindrical coordinate system for spheres
(taking advantage of the axial symmetry). In both cases, an overall finite-size box is of a rectangular shape of dimensions $L_x\times L_z$ with $L_x$ chosen up to
$500\,\sigma$ and $L_z$ up to $200\,\sigma$. By determining the equilibrium density profile, the excess (over ideal) grand potential $\Omega_{\rm ex}$ can be calculated.
The effective potential between the pair of immersed particles a distance $H$ apart is given by
 \bb
 W(H)=\Omega_{\rm ex}(H)-2\Omega_{\rm ex}^{(1)}\,,
 \ee
where  $\Omega_{\rm ex}^{(1)}$ is the excess grand potential corresponding to a system where only one colloidal particle is present. The derivative $f_s\equiv-\dd W/\dd
H$ defines an effective force of solvation between the spheres or cylinders. Using this microscopic formalism we can go beyond macroscopics and classify the order of
bridging transitions. As the sphere/cylinder separation is reduced a first-order bridging may occur at $H=H_b$, where the value $H_b=H_b(R)$ is a function of the
sphere/cylinder radius. For large $R$ we anticipate that this $H_b(R)$ is a near linear function of $R$ in accord with macroscopic expectations. At the separation $H_b$
the potential $W(H)$ has a kink meaning that the solvation force $f_s(H)$ is discontinuous and exhibits a finite jump at $H=H_b$.  It is, however, conceivable that the
line of first-order bridging transitions $H_b(R)$ terminates at a critical point $R=R_c$ at which  the solvation force is infinitely steep, i.e. $\dd f_s/\dd H=\infty$.
The bridging transition is then critical. Finally, for $R<R_c$ the solvation force $f_s(H)$ is a continuous function from which one can identify a rounded bridging
transition at $H=H_b$ where  the curvature of the solvation force changes sign as given by $\dd^2f_s/\dd H^2=0$. The possible nature of bridging transitions are
illustrated in Fig.~\ref{sketch_solv}.

Our hard-wall colloids, which are completely dry, are immersed in saturated liquid (solvent) with the temperature set to $T=0.89\,T_c$ where $k_BT_c=0.9\,\varepsilon$ is
the bulk critical temperature. For each value of the colloidal separation $H$ and different radii $R$ we minimize the grand potential functional $\Omega[\rhor]$ on a
two-dimensional grid to obtain the equilibrium density profile. This is done by starting from two different initial configurations corresponding to a bridged and an
unbridged state where the bridge is formed by a single film of a low-density gas connecting the colloids. In standard fashion, convergence to distinct configurations
then indicates the presence of stable and metastable states and the crossing of stable branches in the $\Omega$-$H$ diagram determines $H_{\rm b}$ for first-order
bridging transitions. The absence of distinct branches indicate that the formation of a bridge as $H$ is continuous corresponding to either critical or rounded
transitions.

\begin{figure}
\includegraphics[width=0.5\textwidth]{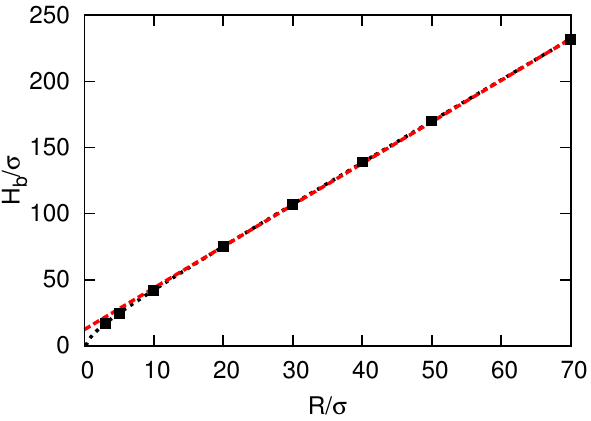}
\caption{(Color online) Phase diagram for cylindrical bridging transitions obtained for $T=0.89\,T_c$ at a two-phase coexistence. The symbols represent DFT results
obtained from a numerical solution of Eq.~(\ref{el2}). The red dashed line is a straight line with a slope of $\pi$ in accord with the macroscopic prediction given by
Eq.~(\ref{kelvin_cyl}). The black dotted line is a fit containing a log correction due to complete wetting by gas around the cylinders parameterized results as $H_b=\pi
R+a\ln R/\sigma$. We find with $a=3.34\,\sigma$ in good agreement with the predicted value $a=\pi\xi_g\approx 3.52$.} \label{pd_cyl}
\end{figure}

\begin{figure}
\includegraphics[width=0.45\textwidth]{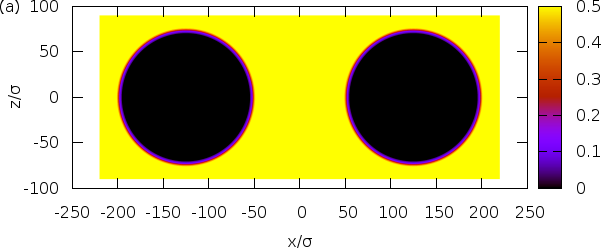} \hspace*{1cm} \includegraphics[width=0.45\textwidth]{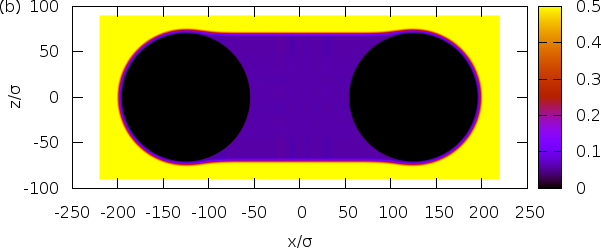}
\caption{(Color online) Two-dimensional density profiles of the saturated liquid near to two infinite cylinders each of a radius $R_c=70\,\sigma$ whose centers are
separated by a distance $H=230\,\sigma$. The density profiles correspond to a first-order bridging transition at which the unbridged state (a) and the bridged state (b)
coexist. } \label{p_cyl}
\end{figure}

\begin{figure}
\includegraphics[width=0.45\textwidth]{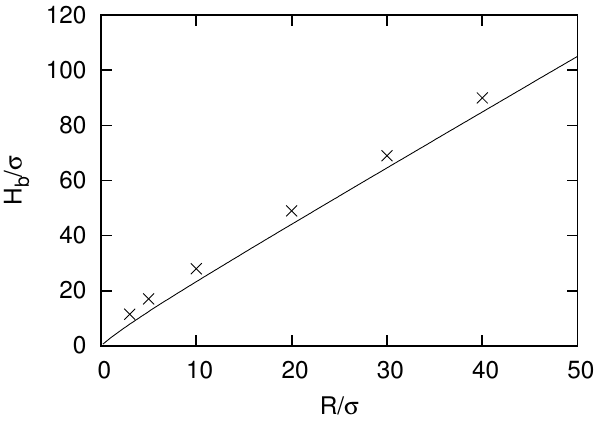}
\caption{Phase diagram for cylindrical bridging transitions obtained for $T=0.89\,T_c$ off coexistence showing comparison between DFT (symbols) and macroscopic theory
(solid line) as given by Eqs.~(\ref{kelvin_cyl_off}) and (\ref{kelvin_cyl_off2}). Here, the Laplace radius $L\approx5\,\sigma$. The bridging transition remains
first-order down to microscopic radii but exhibits a critical point below the radius $R=3\,\sigma$ (see the text).  Note that in contrast to the case of coexistence
shown in Fig.~\ref{pd_cyl} where the macroscopic theory was confirmed by analyzing the slope of the phase boundary for sufficiently large $R$, here the full comparison
is made for the whole range of radii considered.}\label{off_coex}
\end{figure}

\begin{figure}
\includegraphics[width=0.35\textwidth]{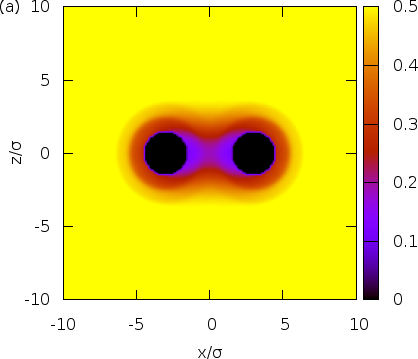} \hspace*{1cm} \includegraphics[width=0.35\textwidth]{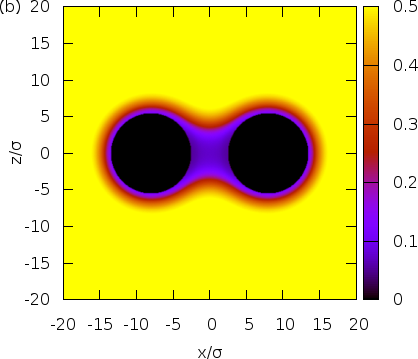}
\caption{(Color online) Two-dimensional density profiles of the supersaturated liquid near to two cylindrical colloids each of a radius $R=\sigma$ (a) and $R=5\,\sigma$
(b) corresponding to the bridging transition. For $R=5\,\sigma$ the transition is first-order and the boundary of the bridging film is near circular with a radius of
curvature $L\approx5\,\sigma$ in line with the macroscopic considerations. For $R=\sigma$ the transition from a bridged to an unbridged state is continuous.}
\label{p_cyl_off}
\end{figure}

In Fig.~\ref{pd_cyl} we show the bridging phase diagram for the values of $H_b$ vs $R$ obtained for cylinders with radii ranging from $R=3\,\sigma$ to $R=70\,\sigma$.
There is excellent agreement with the macroscopic prediction $H_b=\pi R$ for $R>10\,\sigma$ and even better agreement over the entire range of $R$ values when we include
a logarithmic correction as expected from the influence of the complete drying layers.  It is surprising that the latter expression provides almost a perfect fit to our
DFT data down to radii of a few of $\sigma$ units, despite the mesoscopic nature of the prediction.  Fig.~\ref{p_cyl} shows two coexisting density profiles corresponding
to bridged and unbridged states for $R=70\,\sigma$. Notice that the liquid-gas interfaces of the gas bridge connecting the cylinders are near planar in agreement with
macroscopic expectations. A remarkable feature is that for the present temperature  a first-order bridging transition occurs down even to the smallest radius considered.
% It may be that if $R$ was reduced further then, at a mean-field level which ignores fluctuation effects, we would eventually encounter a critical radius
%$R_c$ below which there is only a unique density profile for all values of the separation $H$. However at these microscopic scales the very concept of bridging loses its
%meaning and also even our microscopic DFT is unreliable since it is unable to capture the fluctuation effects that dominate at these distances. We return to this in our
%discussion.
We have also determined bridging phase boundary off bulk coexistence, results of which are shown in Fig.~\ref{off_coex}. In contrast to the case of bulk coexistence, the
$H_b$-$R$ dependence is no more linear but rather follows the macroscopic law given by Eq.~(\ref{kelvin_cyl_off}) with two linear regimes appropriate for small and large
$R$. The bridging transition remains first-order down to about $R_c\approx3\,\sigma$, but exhibits continuous (rounded) transition below. In Fig.~\ref{p_cyl_off} we show
two density profiles corresponding to the phase boundary for $R=5\,\sigma$ (first-order bridging, bridged state) and $R=\sigma$ (rounded bridging).

\begin{figure}
\includegraphics[width=0.5\textwidth]{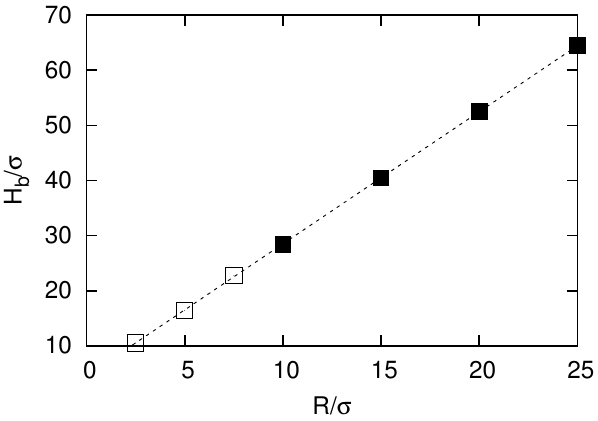}
\caption{Phase diagram for spherical bridging transitions obtained for $T=0.89\,T_c$. The symbols refer to the DFT results: the filled squares denote points for which
the bridging transition is first-order, while the empty squares correspond to a continuous bridging transition. The line represents a linear fit with slope $2.38$ close
to the predicted macroscopic value $2.32$. The critical radius is estimated as $R_c\approx 10\,\sigma$.} \label{pd_spheres}
\end{figure}

\begin{figure}
\includegraphics[width=0.45\textwidth]{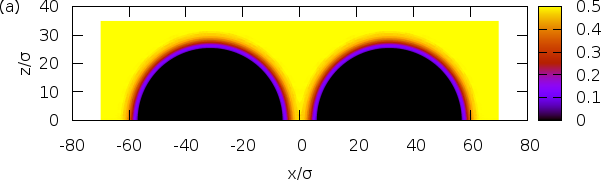} \hspace*{1cm} \includegraphics[width=0.45\textwidth]{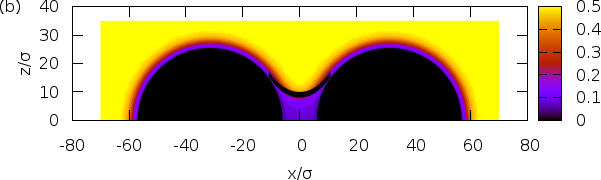}
\caption{(Color online) Two-dimensional density profiles of the saturated liquid near to two spherical colloids each of a radius $R=25\,\sigma$ whose centers are
separated by a distance $H=64\,\sigma$. The density profiles correspond to a first-order bridging transition at which the unbridged state (a) and the bridged state (b)
coexist. A fit to a catenoid shape is shown (black line).} \label{p_spheres}
\end{figure}

\begin{figure}
\includegraphics[width=0.5\textwidth]{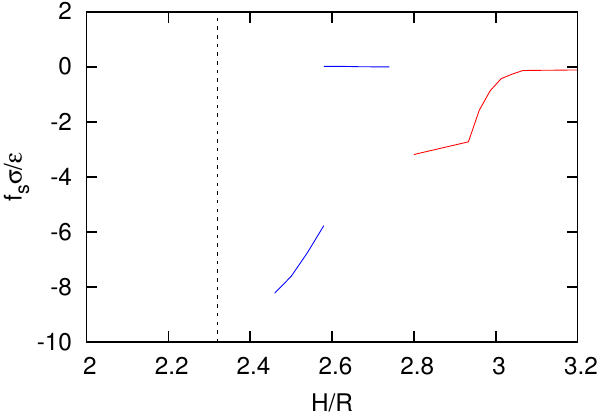}
\caption{(Color online) DFT results for the solvation force $f_s(H)$ of a fluid with short range forces adsorbed between two spheres as a function of separation for two
different radii. (a) $R=25\,\sigma>R_c$, first-order bridging (blue line), (b) $R=7.5\,\sigma<R_c$, rounded bridging (red line, rightmost). The dotted vertical line
represents the macroscopic prediction for $H_b/R=2.32$. } \label{sph_solv}
\end{figure}

For spheres, the phase diagram is similar but the role of a critical radius $R_c$ is more pronounced presumably because of the weakened stability of the bridging film.
This is illustrated in Fig.~\ref{pd_spheres} showing the phase diagram obtained for the same temperature $T=0.89\,T_c$ with radii ranging now from $R=3\,\sigma$ to
$R=25\,\sigma$. In this case we find that a first-order bridging transition only occurs for $R>R_c$ with $R_c\approx 10\,\sigma$. In the first-order regime there is once
again near perfect agreement with the macroscopic prediction $H_b=2.32\,R$. Logarithmic deviations from this are not at all significant as the radius $R$ is large. Also
recall that the anticipated pre-factor of any log correction is only half of that predicted for cylinders. Coexisting density profiles and the fit to a catenoid are
shown in Fig.~\ref{p_spheres} . For $R<R_c$, as the separation $H$ is reduced, a gaseous bridge forms continuously between the spheres without the occurrence of any
singular behavior in  the effective potential. Thus the solvation force $f_s(H)$ shows no jump and is a continuous function. However, we may identify a rounded bridging
transition for $R<R_c$ which occurs when there is a point of inflection $f_s''(H)=0$, as shown in Fig.~\ref{sph_solv} together with the solvation force behaviour for
$R>R_c$ exhibiting a finite jump. Remarkably, the macroscopic prediction $H_b=2.32\,R$ extends through the critical point at $R_c$ and predicts rather accurately the
location of the rounded transition (cf. Fig.~\ref{pd_spheres}).

\begin{figure}
\includegraphics[width=0.5\textwidth]{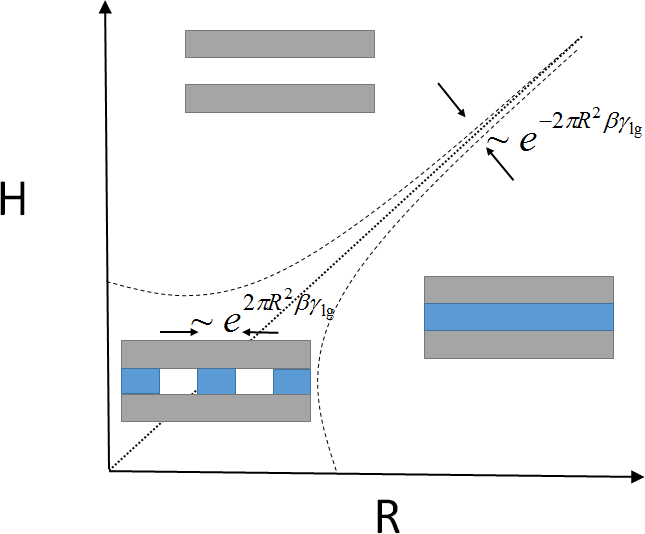}
\caption{(Color online) Schematic phase diagram for bridging transitions and typical configurations between completely wet (dry) cylinders taking into account the effect
of fluctuations. The dotted line represents the location of the macroscopic phase boundary $H_b=\pi R$ whilst the dashed line represents the cross-over region of width
$\sim e^{-2\pi \beta\gamma_{lg}R^2}$ centered around $H_b(r)$ over which the bridging transition is rounded. In the cross-over region we have highlighted the typical
size of the domains of bridged and unbridged regions. }\label{fluct}
\end{figure}

\section{Discussion}

In summary, we have used macroscopic theory and microscopic DFT to study the bridging transition between cylinders and between spheres. The transitions are both
qualitatively similar to condensation occurring in capillary-slits and pores but are quantitatively different to each other due to the very different stability
properties of the liquid bridge. In particular for spheres, DFT studies show that there is always a critical radius $R_c$ below which the bridging is rounded while for
cylinders, the limit of stability depends on the Laplace radius of the bridging interface, so that the transition remains first-order at bulk coexistence and away from
the critical temperature. This echoes the stability properties of the liquid bridge within the macroscopic theory where it is only for spheres that, at two phase
coexistence, the spinodal and coexistence lines lie close to each other. For bridging between spheres our DFT studies indicate that the critical radius $R_c$ increases
with temperature. For example repeating our numerical analysis we have found that at $T=0.94\,T_c$ the value of the critical radius for spheres is
$R_c\approx30\,\sigma$. Indeed, for a near critical temperature $T=0.98\,T_c$ we could not detect first-order bridging transition at all for any values of the colloids
radii up to $R=70\,\sigma$. This is consistent with expectations of finite-size scaling that $R_c$ should increase with $\xi_l$ or $\xi_g$. Indeed, for spheres recall
that even at the macroscopic level the distance between the bridging transition and its spinodal at bulk coexistence, is extremely small with,  $H_{\rm spin}-H_b\approx
0.06 R$. Minor corrections to either of the lines, $H_b(R)$ and/or $H_{\rm spin}(R)$, coming from microscopic length-scales, may cause these to cross, which would
indicate the location of a critical radius. A natural candidate for such a microscopic length is the wetting layer thickness which is $\propto \xi_l\ln R/\xi_l$, as
predicted by Eq.~(\ref{cyl_derj}). Assuming that $H_{\rm spin}(R_c)-H_b(R_c)\propto  \xi_l\ln R_c/\xi_l$ yields a value for $R_c$ which scales with the bulk correlation
length but with a large pre-factor.

The role of a critical radius $R_c$ is at least quantitatively different for the case of cylinders. In this case we are able to find first-order bridging transitions
that persist much closer to $T_c$ and down to small radii. This is consistent with the above finite-size scaling argument since at bulk coexistence there is no
macroscopic spinodal line. However, discussions concerning the existence of a critical radius $R_c$ and its behaviour as $T$ both for spheres and cylinders increases are
somewhat academic, since beyond the mean-field, fluctuation effects destroy all true phase equilibria, so that all bridging transitions are rounded. For bridging between
spheres the volume available to the gas or liquid bridge is finite, meaning that the first-order boundary $H_b(R)$ must be smeared over a region  proportional to the
volume of the system \cite{finite}, i.e. of order $R^{-3}$. Similarly for cylinders, where the bridge extends along the cylinders, we anticipate that the rounding is
sharper and is over order ${\rm e}^{-2\pi \beta\gamma_{\rm lg}R^2}$ which is Boltzmann's factor corresponding to the probability to form a liquid-gas domain, since the
geometry is pseudo one-dimensional. In this case, near to the macroscopic line $H_b=\pi R$ the bridge breaks up into domains of bridged and unbridged states of extension
$\xi\approx {\rm e}^{2\pi\beta\gamma_{\rm lg}R^2}$, similar to the pseudo-symmetry breaking in 2D Ising strips, see Fig.~\ref{fluct}. It is this rounding which provides
the true mechanism, beyond mean-field, for the disappearance of even rounded bridging transitions as the temperature is increased to $T_c$.

In this work we restricted our microscopic (DFT) study for the models in which the fluid particles interact purely repulsively with the walls (cylinders or spheres).
Although from a macroscopic perspective the situation in which the fluid interacts attractively with the walls and a bulk phase is vapour rather than gas is completely
analogous, one expects  a richer phase behaviour of such systems with prewetting and layering phenomena also involved when a more microscopic treatment is used.
Moreover, the high-density phase near the walls will exhibit much stronger excluded volume effects (not included within the macro- or mesoscopic approaches) so that the
comparison between analytic predictions and numerical DFT results is expected to reveal somewhat larger deviations.

It would also be interesting to consider other geometries such as crossed or non-parallel cylinders for which the shape of the bridging film is more complicated
requiring a three dimensional DFT analysis. In this case we anticipate that the bridge may undergo further phase transitions and associated interfacial fluctuation
effects. An example of this would be reducing the opening angle between two slightly misaligned, near parallel, cylinders, in which case the length of the bridging film
must diverge. This is similar, but not identical to filling transition for fluids adsorbed in wedge geometries.

\begin{acknowledgments}
 \noindent A.M. acknowledges the support from the Czech Science Foundation, project 13-09914S. A.O.P. wishes to thank the support of the EPSRC UK
 for Grant No. EP/J009636/1.
\end{acknowledgments}


\begin{thebibliography}{99}

\bibitem{lek}
H. N. W. Lekkerkerker and R. Tuinier, {\it Colloids and the Depletion Interaction}, Springer Netherlands (2011).

%\bibitem{dijk}
% M. Dijkstra, R. Van Roij, and R. Evans, Phys. Rev. Lett {\bf 81}, 2268 (1998).

%\bibitem{roth}
%R. Roth, R. Evans, and S. Dietrich, Phys. Rev. E {\bf 62} 5360 (2000).

\bibitem{macro_rings}
F. M. Orr, L. E. Scriven, and A. P. Rivas,  J . Fluid Mech. {\bf 67}, 723 (1975).

\bibitem{macro_vogel}
T. I. Vogel, Pacific J. Math. {\bf 224}, 367 (2006).

\bibitem{macro_willet}
C. D. Willett, M. J. Adams, S. A. Johnson, and J. P. K. Seville, Langmuir {\bf 16}, 9396 (2000).

\bibitem{yeomans}
H. T. Dobbs, G. A. Darbellay, and J. M. Yeomans, Europhys. Lett. {\bf 18}, 439 (1992).

\bibitem{yeomans2}
H. T. Dobbs and J. M. Yeomans, J. Phys.: Condens. Matter {\bf 4}, 10133 (1992).


\bibitem{dietrich}
C. Bauer, T. Bieker, and S. Dietrich, Phys. Rev. E {\bf 62}, 5324 (2000).

 \bibitem{archer1}
 A. J. Archer, R. Evans, R. Roth, and M. Oettel, J. Chem. Phys. {\bf 122}, 084513 (2005).

\bibitem{archer2}
 P. Hopkins, A. J. Archer, and R. Evans, J. Chem. Phys, {\bf 131}, 124704 (2009).

\bibitem{mal}
A. Malijevsk\'y, Mol. Phys. {\bf 113}, 1170 (2015).

\bibitem{macro_cyl}
A. E. S\'aez and R. G. Carbonell,  J . Fluid Mech. {\bf 176}, 357 (1987).

\bibitem{gelfand}
M. P. Gelfand  and R. Lipowsky, Phys. Rev. B {\bf 36} 8725 (1987).

\bibitem{bieker}
T. Bieker and S. Dietrich, Physica A {\bf 252}, 85 (1998).

\bibitem{stewart}
M. C. Stewart and R. Evans, Phys. Rev. E {\bf 71}, 011602 (2005).

\bibitem{nold}
A. Nold, A. Malijevsk\'y, and S. Kalliadasis, Phys. Rev. E {\bf 84}, 021603 (2011).



\bibitem{evans79}
R. Evans, Adv. Phys. A {\bf 28}, 143 (1979).


\bibitem{rosenfeld89}
Y. Rosenfeld, Phys. Rev. Lett. {\bf 63}, 980 (1989).

\bibitem{mal_grooves}
A. Malijevsk\'y, J. Phys.: Condens. Matter {\bf 25}, 445006 (2013).

\bibitem{mulero}
P. Tarazona, J. A. Cuesta, and Y. Martinez-Raton, in A. Mulero (Ed.), Theory and Simulation of Hard-Sphere Fluids and Related Systems, Lect. Notes Phys. 753 (Springer,
Berlin Heidelberg 2008), pp. 247-382.

\bibitem{our_wedge}
A. Malijevsk\'y and A. O. Parry, J. Phys.: Condens. Matter {\bf 25}, 305005 (2013).

\bibitem{finite}
V. Privman and M. E. Fisher, J. Stat. Phys. {\bf 33}, 385 (1983).


\end{thebibliography}
\end{document}